# Normal Coordinates Describing Coupled Oscillations in the Gravitational Field

Walter James Christensen Jr.
Department of Physics Cal State University, Fullerton
800 N. State College Blvd. Fullerton, CA 92831
And
Department of Physics Cal Poly Pomona University
3801 W. Temple Blvd. Pomona, CA 91768

## Abstract

The motion of a local source inducing small oscillations in the gravitational field is investigated and shown to exhibit pure rotational kinetic energy. Should the net affect of these slow, revolving oscillations cause large-scale rotations in spacetime it would certainly result in anomalous celestial accelerations. When this angular rotational frequency of spacetime is applied to the anomalous acceleration of the Pioneer 10/11 spacecrafts, the correlation is promising.

## I. Introduction

Fundamental to the theory of general relativity is the coupling existing between the gravitational field and the energy-momentum source $T_{\mu\nu}$; if one changes, so too, will the other. In particular, if the gravitational field undergoes oscillations then there must be a causal source inducing these oscillations. If so, this suggests the gravitational system can be treated like a coupled spring and driver. Though coupled motion can be quite complex, not even periodic, it can always be described in terms of a set of normal coordinates having the property that each coordinate oscillates with a single, well-defined frequency with no coupling among them[1].

The goal then will be to describe the motion of the energy-momentum-source $T_{\mu\nu}$, simply by knowing that the gravitational field is oscillating. This can be accomplished analogously through the classical approach of analyzing small displacements about a point of equilibrium and then solving for the normal coordinates, a procedure that is also well known in gravitational literature[2]. Once these coordinates are identified they can be brought into the language of general relativity. The metric tensor is then constructed and energy-momentum tensor calculated from the Einstein Tensor $G_{\mu\nu}$. The resulting diagonal tensor $T_{\mu\nu}$ will be shown to have components of pure rotational kinetic energy density. In classical physics this diagonal-kinetic-energy result is a necessary condition imposed by normal coordinates. Therefore the method presented here of extending normal coordinates into general relativity is promising. Furthermore, though the diagonal rank-two tensor $T_{\mu\nu}$ is shown to be constant and real, the contravariant tensor $T^{\mu\nu}$ necessarily turns out to be complex. However, $T^{\mu\nu}$ becomes real and equal to $T_{\mu\nu}$ every one-fourth the period of the fundamental mode of oscillation of the normal coordinates. Countable rotational symmetry, together with Noether's theorem[3], suggests the energy-momentum tensor $T^{\mu\nu}$ is conserved.

## II. Normal Coordinates

Einstein's gravitational field equations express a causal link between the energy-momentum-source $T_{\mu\nu}$, and spacetime curvature associated with the tensor $G_{\mu\nu}$. The purpose of this paper will be to determine the causal motion of the source inducing oscillations in the gravitational field. The problem is reminiscent of a classical oscillator and driver and will be our starting point. We begin with a Lagrangian representing small oscillations about a point of equilibrium.

$$L = \frac{1}{2}\left(T_{ij}\dot{\eta}_i\dot{\eta}_j - V_{ij}\eta_i\eta_j\right) \tag{2.1}$$

The $\eta_i$'s represent small deviations from the generalized coordinates $q_{0i}$, such that $q_i = q_{0i} + \eta_i$. Classically the $\eta$'s subsequently become the generalized coordinates for the equations of motion, wherein the kinetic energy has diagonal components only.

$$T_i\ddot{\eta}_i - V_{ij}\eta_j = 0 \qquad \text{(no sum over i)} \tag{2.2}$$

The solution[4] to (2.2) has the normal coordinate form of

$$\eta_i = C_\kappa e^{-i\omega_\kappa t} \tag{2.3}$$

Assuming these coordinates quasi-describe oscillations in the gravitational field, by the principle of equivalence let a general relativistic coordinate basis $e_\mu$ experience the accelerations expressed by (2.3). Furthermore let the coefficients of (2.3) be set equal to one and let negative one-half be introduced in front of the angular velocity. These small changes allow for the motion of the energy-momentum source to become more apparent. Rayleigh's principle[5, 6] is applied, and the coordinate frequencies $\omega_\kappa$ reduce to the fundamental mode of oscillation, $\omega$, having the greatest intensity. The average kinetic energy $\langle T \rangle$ is then equal to the average potential energy $\langle U \rangle$.

With the preceding adjustments made, the basis for the general relativistic coordinate system is constructed from the modified normal coordinates[7, 8]:

$$\left(e_\mu\right)_\nu \equiv e^{\frac{i\omega t}{2}} \delta_{\mu\nu} \tag{2.4}$$

By definition the inner product of any two such basis elements $e_\mu, e_\nu$ yields the metric tensor.

$$g_{\mu\nu} \equiv e_\mu \cdot e_\nu = e^{i\omega t} \delta^\mu{}_\nu = e^{i\omega t} \eta_{\mu\nu} \tag{2.5}$$

As with the mechanical oscillation problems it is understood that only the real part of this complex metric corresponds to physical measurement.

### III. Energy-Momentum Tensor

The constructed weak field metric $g_{\mu\nu}$ is applied to Einstein tensor $G_{\mu\nu}$. A straightforward calculation produces the energy-momentum tensor $T_{\mu\nu}$[9], and together they form a linearized theory of gravitation:

$$T_{\mu\nu} = \frac{c^2}{16\pi G} \begin{bmatrix} -\frac{3\omega^2}{2} & 0 & 0 & 0 \\ 0 & \frac{\omega^2}{2} & 0 & 0 \\ 0 & 0 & \frac{\omega^2}{2} & 0 \\ 0 & 0 & 0 & \frac{\omega^2}{2} \end{bmatrix} \quad (3.1)$$

$T_{\mu\nu}$ is separated out to show its rotational kinetic energy density form.

$$T_{\mu\nu} = \frac{1}{2}\frac{c^2}{8\pi G} \begin{bmatrix} -\frac{3}{2} & 0 & 0 & 0 \\ 0 & \frac{1}{2} & 0 & 0 \\ 0 & 0 & \frac{1}{2} & 0 \\ 0 & 0 & 0 & \frac{1}{2} \end{bmatrix} \begin{bmatrix} \omega^2 & 0 & 0 & 0 \\ 0 & \omega^2 & 0 & 0 \\ 0 & 0 & \omega^2 & 0 \\ 0 & 0 & 0 & \omega^2 \end{bmatrix} \quad (3.2)$$

The moment of inertia and angular frequency matrices are defined to be

$$I \equiv \frac{c^2}{8\pi G} \begin{bmatrix} -\frac{3}{2} & 0 & 0 & 0 \\ 0 & \frac{1}{2} & 0 & 0 \\ 0 & 0 & \frac{1}{2} & 0 \\ 0 & 0 & 0 & \frac{1}{2} \end{bmatrix} ; \quad \tilde{\omega}^2 \equiv \begin{bmatrix} \omega^2 & 0 & 0 & 0 \\ 0 & \omega^2 & 0 & 0 \\ 0 & 0 & \omega^2 & 0 \\ 0 & 0 & 0 & \omega^2 \end{bmatrix} \quad (3.3)$$

In compact tensor notation the energy-momentum tensor becomes

$$T_{\mu\nu} \equiv \frac{1}{2}I\tilde{\omega}^2 \qquad (3.4)$$

If the angular velocity is replaced by $\frac{v}{r} = \omega$, then equation (3.1) resembles an energy-momentum tensor for a radiation dominated perfect fluid[10]--in particular a perfect fluid of gravitons. It is important to realize $T_{\mu\nu}$ was derived from Einstein's gravitational wave equation based on a variational principle, and not upon the prejudice of definition. Furthermore, it is interesting to observe that, although, $T_{\mu\nu}$ is completely real, its contravariant counterpart $T^{\mu\nu}$ is necessarily complex.

$$T^{\mu\nu} = g^{\mu\alpha}g^{\nu\beta}T_{\alpha\beta} = e^{-2i\omega t}\eta^{\mu\alpha}\eta^{\nu\beta}T_{\alpha\beta} \qquad (3.5)$$

This result shows the Einstein tensor $G_{\mu\nu}$ and its metric constructed from normal coordinates, are able to separate the energy momentum-tensor $T^{\mu\nu}$ into real and imaginary parts through a time rotation. Although time-wise there are uncountable many complex energy-momentum tensors, $T^{\mu\nu}$ becomes completely real and equal to $T_{\mu\nu}$ every one-fourth the period of the fundamental mode of oscillation; that is whenever $t = \frac{nT}{4}$. Countable symmetry together with Noether's Theorem suggests the energy momentum tensor is conserved under time rotation.

## IV. Estimate of the Fundamental Mode of Oscillation

The preceding developments for an oscillating gravitational system are now applied to a region of spacetime below the micro-level. In this ultra small region both the source and field must be comprised of nearly the same particle, otherwise the source would overdrive the gravitational system and no longer would small oscillations occur about the point of equilibrium. In the world of particle physics the gravitational field is made up of self-interacting gravitons, which also interact with every other particle in the universe. Since the graviton has an extremely small rest mass of less than $2 \times 10^{-65}$ Kg[11], and that the region of spacetime being considered is so very tiny, the proposed source and gravitational field must be comprised of coupled gravitons that oscillate with rotational kinetic energy. These gravitons must therefore have an associated spin $I\tilde{\omega}$. The discrete energy components are notably independent of Planck's constant.

The model envisioned in this tiny region of spacetime is a gyroscopic graviton creating a point of equilibrium from which coupled gravitons not only oscillate at the fundamental frequency $\omega$, they rotate at this frequency as well. Furthermore, since the region is so small, accordingly this frequency must be the de Broglie wave-frequency for a graviton. If such a coupled system is prevalent throughout the universe, the net affect could add to cause large-scale rotations in spacetime. Depending on where an observer resides relative to the axis of rotation, one might see distant bodies exhibiting anomalous accelerations. This assumption is supported by recent radiometric data received from

Pioneer 10/11 spacecrafts, wherein an anomalous inbound acceleration toward the sun is observed. Presently there is no conclusive explanation for this phenomenon.

As a physical check to determine if the anomalous acceleration is related to the rotational frequency of coupled gravitons, the simple calculation $a_r = R\omega^2$ is made for the large-scale rotation frequency $\omega$, of spacetime. The observed inward solar acceleration is $a_r = 8.74 \times 10^{-10} \frac{m}{s^2}$ [12,13]. The distance of 20 AU is chosen for the radial distance R from the axis of rotation to the spacecraft because it is the approximate distance when the Pioneer anomaly was first discovered. The angular velocity of the gravitational field, and hence the de Broglie graviton wave frequency, is computed to be

$$\omega \cong 1.7 \times 10^{-11} \sec^{-1} \qquad (4.1)$$

The de Broglie graviton wave-frequency computed from the graviton mass is

$$\omega_g = \frac{mc^2}{\hbar} = 1.7 \times 10^{-14} \sec^{-1} \qquad (4.2)$$

Though these frequencies are three orders of magnitude apart, as a first approximation the result is promising, especially realizing other values for the graviton mass[14] are nearly six orders of magnitude heavier than the calculated Goldhaber and Nieto result.

## V. Conclusion

In this paper normal coordinates were applied to an oscillating gravitational field as a method for determining the motion of the energy-momentum-source. This motion was calculated to be pure rotational kinetic energy. By applying this result to a region of spacetime below the sub-micro level, the particle's rotation becomes graviton spin. The net affect of this gyroscopic motion has a cumulative affect causing spacetime to rotate at the graviton frequency $\omega$. Calculations based on the pioneer spacecraft data support this notion.

## Acknowledgements

My warmest thanks to Alfonso Agnew for his enthusiastic support, guidance, and for verifying my energy-momentum result is correct. My friendship goes to John Fang, my early mentor on gravitational particle physics. I also wish to extend my friendly thanks (for many reasons) to Kai Lam, Mary Mogge and Heidi Fearn. Lastly, I would like to thank Kip Thorne for spending time with me at the 22$^{nd}$ Pacific Coast Gravitational Meeting to answer questions related to this paper.

[1] The German mathematician Karl Weierstrass (1815-1897) showed in 1858 that the motion of a dynamical system could always be expressed in terms of normal coordinates.

[2] W. Misner, K. S. Thorne and J. Wheeler. Gravitation. W. H. Freeman and Company. 1973. Pages 1022-1034

[3] N. Byers. Israel Mathematical Conference Proceedings Vol. 12, (1999)

[4] H. Goldstein, C. Poole, J. Safko. Classical Mechanics, Addison Wesley. 250 (2002)

[5] J. B. Marion and S. T. Thornton. Classical Dynamics of Particles and Systems, HBJ. 458 (1988, 3rd edition)

[6] R. M. Quick and H. G. Miller. Phys. Rev. D **31**, 2682 (1985)

[7] A. Majid, A. Allezy, R. Dufour. Journal of Vibrations and Acoustics. **128**, 50 (2006)

[8] H. Collins, B. Holdom. Phys. Rev. D **65**, 124014-1 (2002)

[9] Private Communication with Prof. Alfonso Agnew, of California State University Fullerton, Mathematics Department independently confirmed the energy-momentum tensor result of equation (3.1)

[10] W. Misner, K. S. Thorne and J. Wheeler. Gravitation. W. H. Freeman and Company. 1973. §22.3

[11] A. S. Goldhaber and M. M. Nieto. Phys. Rev. D **9**, 1119 (1974).

[12] J. D. Anderson, P. A. Laing, E. L. Lau, A. S. Liu, M. M. Nieto and S. G. Turyshev, Phys. Rev. D **65**, 082004 (2002)

[13] O. Bertolami and J. Parámos. Phys. Rev. D **71**, 023521-3 (2005).

[14] A. Cooray and N. Seto. Phys. Rev. D **69**, 103502 (2004)